\documentclass[prb,aps,twocolumn,superscriptaddress,10pt,showpacs,longbibliography]{revtex4-1}
\usepackage{graphicx}
\usepackage{epstopdf}
\usepackage{color,soul}
\usepackage{amsmath,amssymb}
\usepackage{fixmath}

\newcommand{\angstrom}{\mbox{\normalfont\AA}}
\usepackage [english]{babel}
\usepackage [autostyle, english = american]{csquotes}
\MakeOuterQuote{"}
\usepackage{textgreek}
\usepackage[export]{adjustbox}
\usepackage{siunitx}

\begin{document}

\title{Recovery of harmonic--like behaviour of the polar mode in BaTiO$_{3}$ at high pressures}

\author{A. Herlihy} 
\affiliation{Department of Chemistry, University of Warwick, Gibbet Hill, Coventry, CV4 7AL, UK}
\affiliation{ISIS Neutron and Muon Facility, Rutherford Appleton Laboratory, Didcot, OX11 0QX, UK}

\author{T. A. Bird}
\affiliation{Department of Chemistry, University of Warwick, Gibbet Hill, Coventry, CV4 7AL, UK}

\author{C. J. Ridley}
\affiliation{ISIS Neutron and Muon Facility, Rutherford Appleton Laboratory, Didcot, OX11 0QX, UK}

\author{C. L. Bull}
\affiliation{ISIS Neutron and Muon Facility, Rutherford Appleton Laboratory, Didcot, OX11 0QX, UK}
\affiliation{School of Chemistry, University of Edinburgh, David Brewster Road, Edinburgh, EH9 3FJ, UK}

\author{N. P. Funnell}
\email{nick.funnell@stfc.ac.uk}
\affiliation{ISIS Neutron and Muon Facility, Rutherford Appleton Laboratory, Didcot, OX11 0QX, UK}

\author{M. S. Senn}
\email{m.senn@warwick.ac.uk}
\affiliation{Department of Chemistry, University of Warwick, Gibbet Hill, Coventry, CV4 7AL, UK}

\date{\today}

\begin{abstract}
The local structure of high pressure BaTiO$_{3}$ has been interrogated by neutron total scattering methods, up to pressures of 4.18 GPa at ambient temperature. Competitive refinements of cubic, tetragonal and rhombohedral distortion modes against pair distribution functions indicate contrasting local structure behaviour of temperature- and pressure-induced cubic BaTiO$_{3}$. Suppression of the mode amplitude, isotropy of the order parameter direction and loss of sensitivity to correlated Ti displacements at high pressure all suggest a high-pressure local structure that is more consistent with the harmonic approximation, rather than an order-disorder model which better describes high-temperature cubic BaTiO$_{3}$ in the vicinity of the tetragonal phase transition.

\end{abstract}

\maketitle

\section{Introduction}

BaTiO$_{3}$ is often given as a classic example of a proper ferroelectric where, due to the second-order Jahn-Teller effect, an off-centering of the Ti$^{4+}$ cation from its TiO$_6$ octahedron results in a net polarisation\cite{Bersuker1966}. The resulting ferroelectric properties and high dielectric constant make BaTiO$_{3}$ a very attractive material for use in devices such as capacitors\cite{Acosta2017}, and the perovskite-structured material (shown in Figure \ref{f1}a) has become the prototypical ferroelectric; intensively studied to understand the link between ferroelectricity and crystal structure. Despite many decades of investigation, there remains an ongoing debate about the nature of the ferroelectric phase transition. Above its Curie temperature (\textit{T}{$_\textup{C}$}), BaTiO$_{3}$ adopts a cubic structure. Below \textit{T}{$_\textup{C}$}, the structure is reduced to a tetragonal symmetry and on decreasing temperature further, BaTiO$_{3}$ transforms to an orthorhombic and, finally, rhombohedral structure\cite{Hippel1946,Megaw1947,Cochran1960,Hayward2002}.

A popular theory, suggested by Cochran et al.\cite{Cochran1960} describes a displacive model whereby Ti$^{4+}$ cations are displaced microscopically along 〈100〉, 〈110〉, and 〈111〉 directions for the tetragonal, orthorhombic and rhombohedral phases respectively. This model however fails to address key observations such as the strong diffuse X-ray scattering in all but the rhombohedral phase\cite{Comes1968,Ravy2007,Pasciak2018} and the presence of first-order Raman excitations in the cubic phase\cite{Quittet1973}. In 1968, Com\'es et al.\cite{Comes1968} proposed an order-disorder (OD) model, also commonly referred to as the `eight-site' model, where the crystallographically-rich phase diagram of BaTiO$_{3}$ is rationalised due to correlations of local Ti displacements along the eight 〈111〉 directions. Correlated displacements of the Ti atom in successive 〈100〉 directions give rise to the observed average symmetry, and it is this underlying disorder that appears to simultaneously reconcile the perceived average symmetry with the anomalous experimental results, discussed above.

Since the first proposal of these two contending models, a multitude of experimental and computational studies have favoured either one of these two possible scenarios. Local probes tend to support an OD model\cite{Ravel1998,Zalar2003,Laulhe2009}, for example, our symmetry-motivated analyses of pair distribution functions (PDFs) of BaTiO$_{3}$ have shown that Ti displacements are rhombohedral-like across all known phases\cite{Senn2016}. However, the observation of heavily-damped modes\cite{Luspin1980,Harada1971,Yamada1969} appears at odds with an OD model, and supports the soft-mode explanation. Furthermore, there is not yet consensus---within the OD interpretation---on the exact nature of the disordered local arrangements of Ti cations, where some reports (\textit{via} solid state NMR\cite{Zalar2003}) suggest a local tetragonal distortion and others support a rhombohedral\cite{Ravel1998,Stern2004} distortion. 

More recently, additional work has come out in support of the soft mode model \cite{Pasciak2018}, where diffuse scattering is attributed to the overdamped anharmonic soft phonon branch. This results in a local probability distribution for the Ti atoms that has a minimum coinciding with the average crystallographic position and a maximum along 〈111〉 directions with an average magnitude of ca.~0.15 \angstrom. It seems that a wealth of experimental and computational observations can either be explained by invoking an OD scenario or considering highly over-damped, anharmonic, soft phonon modes that imply the Ti atoms spend a substantial amount of time off-centre. Regardless of the perspective adopted, it is clear that the local symmetry deviates substantially from the average crystallographic symmetry over short length scales and long time periods, indicating a significant departure from the harmonic soft mode/displacive picture. Consideration of the long range ordering of dynamic 〈111〉 Ti displacements projected onto the 〈100〉 directions appears to reconcile these two models\cite{Senn2016}.

Clearly, the investigation of the temperature-induced phase transitions of BaTiO$_{3}$ has been extensive, and a wide range of techniques have been utilised to investigate the average and local structure of the perovskite material \cite{Culbertson2020}. However, challenges associated with \textit{in situ} high pressure measurements have perhaps limited investigation of the local structure of BaTiO$_{3}$ in other regions of the phase diagram. 

It is predicted that modest hydrostatic pressure will initially act to suppress ferroelectric distortions in ABO$_{3}$ perovskites due to the increasing influence of short-range electronic repulsions over long-range Coulomb ionic interactions which favour polar distortions\cite{Kornev2005}. This is born out by the well-established average structure phase diagram of BaTiO$_{3}$ that indicates that at ambient temperature, there is a tetragonal-to-cubic phase transition at ca.~2 GPa \cite{Hayward2002,Ishidate1997,Bull2021}. However, high-pressure Raman studies show evidence for persistent disorder within the cubic phase, with the suggestion that this disorder results from off-centre Ti atoms and grain boundary/intergrain stress \cite{Venkateswaran1998}. X-ray absorption spectroscopy (XAS) of the Ti \textit{K} edge also suggests that Ti remains displaced until 10 GPa, above which the Ti is centred, and local and average symmetries are reconciled\cite{Itie2006}. Together, these results might imply that the high temperature and high pressure behaviour mimic each other from both an average (crystallographic) and local structure perspective. However, neither of these studies appear to have allowed for robust refinement of models with competing symmetries against the local probe data. 

On the other hand, PDFs generated from total scattering experiments and their sensitivity to short-range atom--atom correlations are well suited to this kind of modelling that interrogates the precise local symmetry breaking behaviour in BaTiO$_{3}$. Whilst X-ray PDF work has been carried out\cite{Ehm2011}, the insensitivity of X-rays to the lighter oxygen atoms often fail to resolve the level of detail available to neutron measurements. The lack of high-pressure neutron PDF studies of BaTiO$_{3}$, and indeed of crystalline materials more generally can be attributed to the often opposing requirements of high pressure and PDF experiments. It is only relatively recently that high-pressure neutron PDF measurements have been achieved for crystalline materials\cite{Playford2017,Herlihy2021}. 

With this in mind, we undertake the first analysis of neutron total scattering measurements of BaTiO$_{3}$ at pressures up to 4.2 GPa, in order to directly investigate the nature of the pressure-induced tetragonal-to-cubic phase transition of BaTiO$_{3}$ over a range of length scales. Building on our recently developed symmetry adapted PDF analysis (SAPA) \cite{Bird2021} technique, whereby distortion modes grouped by irreducible representation are refined against local structure measurements, we analyse the high-pressure PDF data, revealing pressure-induced suppression of the local Ti off-centerings. We apply the same modelling approach to previously-published variable temperature PDFs\cite{Senn2016} in order to determine how the departure of local from average symmetry compares for pressure \textit{vs} temperature. Our analysis of ambient temperature, variable pressure PDFs points toward a gradual pressure-induced suppression of the anharmonic potential implicit in describing the OD behaviour of BaTiO$_{3}$, towards a more harmonic-like potential, more consistent with a soft-mode picture.

\section{Experimental Details and Data Analysis}

Polycrystalline BaTiO$_{3}$ (also used for the variable-temperature study described in reference \citenum{Senn2016}) was measured\cite{DATA} on the high-pressure instrument PEARL\cite{Bull2016}, at the ISIS Neutron and Muon Facility. The powder sample was loaded into a null-scattering Ti--Zr single-toroidal gasket, with a gas loader\cite{Klotz2013}, used to fill the remaining gasket volume with an argon gas pressure transmitting media (PTM). A Paris--Edinburgh (PE) press, equipped with zirconia-toughened alumina (ZTA) anvils was used to apply loads of 3, 25, 40 and 50 tonnes to the sample. The lattice parameters of BaTiO$_{3}$ were determined from Rietveld refinement against the Bragg data and the known equation of state\cite{Bull2021} used to calculate sample pressures of 0.24(2), 1.19(2), 2.55(6) and 4.18(8) GPa. Neutron powder diffraction patterns were collected for a minimum of 11 hours each to ensure sufficient signal-to-noise ratio at high $Q$ (where $Q=(4\pi\sin\theta)/\lambda$). Stacked vanadium discs were measured in the same way with an argon PTM, and analogous data collections were performed at loads of 8, 20, 30 and 45 tonnes, corresponding to pressures roughly equivalent to those of the measured BaTiO$_{3}$ data. Total scattering data were collected and treated using the same procedure described in references \citenum{Playford2017} and \citenum{Herlihy2021}, without the added complication of needing to model the PTM, since argon gas is a relatively weak neutron scatterer. That being said, scattering due to the PTM was observed, with the presence of the (111) Bragg reflection in the diffraction pattern at 4.18 GPa, suggesting the PTM had crystallised. However, the absence of any significant sample peak broadening indicated that hydrostatic conditions remained and there was no evidence of an argon contribution to the PDF (i.e.~no misfits in regions where an Ar--Ar peak would be expected). 

\begin{figure}[t!]
	\includegraphics[width=0.5\textwidth]{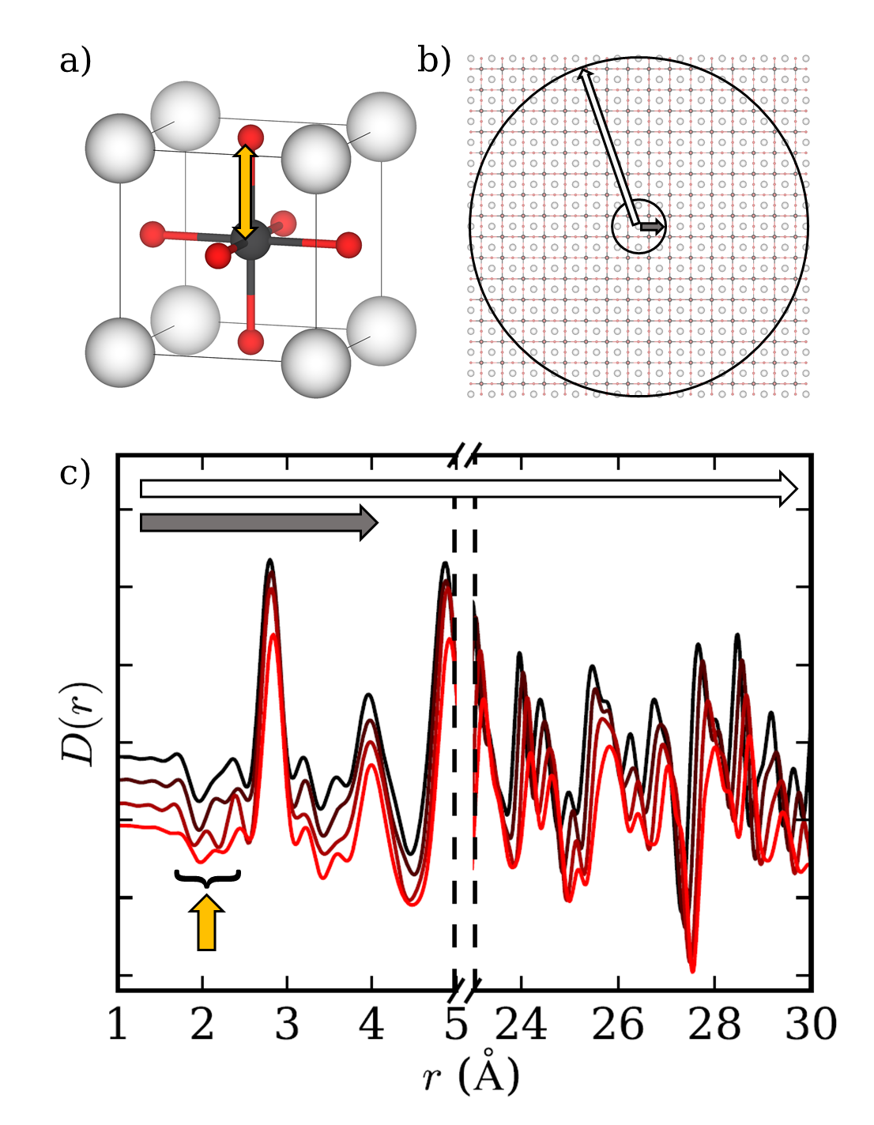}
	\caption{\label{f1} a) The average structure unit cell of cubic BaTiO$_{3}$, with an arrow indicating the shortest atom--atom correlation within the structure (Ti--O). b) The longer-range structure and circles with radii of 4 and 30 \angstrom~indicating the minimum and maximum range of atom--atom correlations probed by our variable range PDF refinements. c) Variable-pressure PDFs measured on PEARL (offset in the y-direction with increasing pressure for clarity). The yellow arrow indicates the features arising due to the Ti--O correlations and the horizontal arrows correspond to the probe distances shown in Figure 1b.}
\end{figure}

Data were reduced using the MANTID software package \cite{Arnold2014} to correct for the effects of attenuation by the ZTA anvils and normalised by a vanadium standard to account for flux profile and detector efficiencies. Scattering from the gasket and anvils were accounted for by subtracting data from the vanadium measurements, and total scattering patterns (\textit{S}(\textit{Q})s) were produced by applying a scaling and offset value such that \textit{S}(\textit{Q})\textrightarrow1 at \textit{Q}\textsubscript{max}. PDFs (shown in Figure \ref{f1}c) were obtained \textit{via} Fourier transform of the \textit{S}(\textit{Q}) function using the program StoG, distributed with the RMCProfile package \cite{Tucker2007}.

PDF modelling and Rietveld refinements were carried out using TOPAS Academic software v6\cite{Coelho2018}. We performed small-box variable range PDF refinements\cite{Smith2008,Culbertson2020}, with the minimum of the fitting range ($r_{\textup{min}}$) kept constant at 1.2 Å, and the maximum ($r_{\textup{max}}$) varied from from 4 to 30 Å in steps of 1 Å. Therefore, the overall fitting range was varied between 2.8 and 28.8 Å, such that increasingly large length-scale atom--atom correlations were probed with increasingly large $r_{\textup{max}}$ values, as depicted in Figure \ref{f1}b. This is in contrast to so-called `box-car' refinements\cite{Usher2016,Hou2018} where the fitting range is held constant and shifted along the PDF, resulting in the progressively reduced influence of the immediate local structure on the refined small-box model. We used a $P1$ unit cell, refining only the polar distortion modes associated with Ti and O which transform as the Γ$_4^-$ irreducible representation (irrep.), and fixing Ba modes to zero to avoid a floating origin of the unit cell. The most general order parameter direction (OPD) associated with this irrep is three dimensional (a,b,c). The Ti($T_{1u}$), O($A_{2u}$) and O($E_{u}$) modes, that form a basis of this irrep, thus have three branches each, where particular constraints on the branched mode amplitudes correspond to higher symmetry OPDs.

Rather than allowing distortion modes to refine freely, we constrained the OPD to be consistent with cubic (0,0,0), tetragonal (a,0,0) and rhombohedral (a,a,a) symmetries in order to test these three specific local distortion behaviours. We did not consider other order parameters such as (a,a,0), (a,b,0) or (a,a,b) as the aim of this work was to resolve the OD behaviour of BaTiO$_{3}$ at the tetragonal to cubic phase transition. We found that unconstrained refinements of the Ti($T_{1u}$), O($A_{2u}$) and O($E_{u}$) modes resulted in non-physical coupling, particularly for refinements of the PDFs measured at 2.55 and 4.18 GPa, where for $r_{\textup{max}}$ values of greater than 10 Å, Ti and O atoms refined to displace in the same, rather than opposite, directions (see Supplementary Information (SI)\cite{SI}). In order to maintain the correct relative displacements associated with the modes, a ratio of 1:$-$1.6:$-$1.3 for Ti($T_{1u}$):O($A_{2u}$):O($E_{u}$) displacements, respectively, was applied. These values were calculated by averaging ratios determined by fitting rhombohedral (a,a,a) models against high quality diffraction data measured at 15 and 293 K on GEM\cite{Senn2016}. Mode amplitude values reported herein refer to the A${_\textup{P}}$ values defined in ISODISTORT\cite{Campbell2006}, as the parent-cell-normalized amplitude, and are assigned the mode-specific notation, $|$Q(Γ$_4^-$)$|$. We found that although the refined $|$Q(Γ$_4^-$)$|$ values differ slightly depending on the precise ratio used, the relative values and fitting statistics of each refinement remain essentially constant. Lattice parameters determined from Rietveld refinements of the diffraction patterns (see SI\cite{SI}) were fixed for all small-box PDF refinements, constraining the metric symmetries to those known from the average structures. The $beq\_r\_r2$ function (discussed further in reference \citenum{Bird2021}) was used to describe the correlated thermal motion that leads to \textit{r}-dependent broadening, with isotropic displacement parameters fixed to the lower limits found for the three models (see SI\cite{SI} for further details). The sensitivity of our modelling approach to the limited {\textit{Q}}\textsubscript{max} (20.32 Å$^{-1}$) available on PEARL was thoroughly investigated and reported within Appendix A.
\\

\section{Results and Discussion}
Neutron diffraction patterns indicate that the measured average structure of BaTiO$_{3}$ at variable pressure is consistent with previous literature\cite{Pruzan2002,Hayward2002,Bull2021}. The neutron diffraction patterns (see SI\cite{SI}) at 0.24 and 1.19 GPa exhibited clear peak splitting (particularly the (200)/(002) reflection), indicative of a tetragonal symmetry, and Rietveld refinements confirmed a $P4mm$ average crystal structure. Above 2 GPa, BaTiO$_{3}$ goes through a well documented phase transition to an average cubic symmetry ($Pm\bar{3}m$), confirmed again by Rietveld refinement at 2.55 and 4.18 GPa.


\begin{figure*}[t!]
	\centering
	\includegraphics[width=\textwidth]{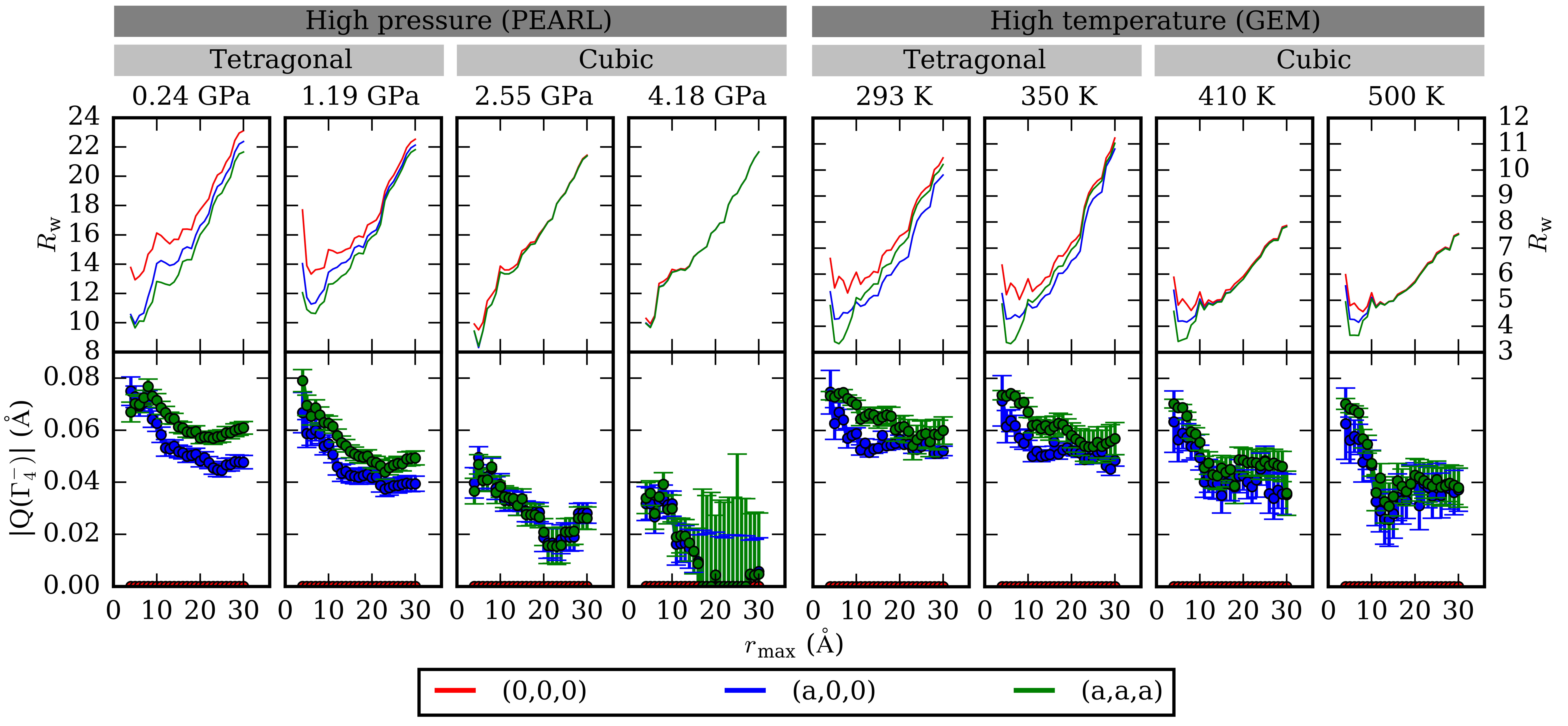}
	\caption{\label{f4} $R_w$ and $|$Q(Γ$_4^-$)$|$ values for variable range refinements for cubic (0,0,0), tetragonal (a,0,0) and rhombohedral (a,a,a) OPDs against variable pressure (left) and temperature (right) PDFs. $|$Q(Γ$_4^-$)$|$ values for the cubic model were fixed at zero and are plotted as such. $R_w$ values for tetragonal and rhombohedral modes in the high pressure cubic data are almost exactly coincident, and cannot be visually distinguished.}
\end{figure*}

Refining small-box models over an increasing range of \textit{r} (Å) of the PDF provides information on the correlation length scale. This is particularly relevant for materials with OD behaviour such as BaTiO$_{3}$ where a local rhombohedral distortion may be observed over a short length scale, for example one unit cell, however, longer length scales will increasingly resemble the average structure. Comparisons of fitting statistics ($R_w$) and $|$Q(Γ$_4^-$)$|$ values (shown in Figure \ref{f4}) for cubic, tetragonal and rhombohedral models provide insight into the evolution of local displacements of Ti and O atoms as a function of pressure. We demonstrate that even for limited \textit{Q}\textsubscript{max} values available on PEARL, our data is sensitive to subtle changes in the local structure (see Appendix A).

We compare our findings for the local structure of BaTiO$_{3}$ at high pressure with analogous results for the thermally-induced phase transition. The same modelling approach has been applied to PDFs measured at 293, 350, 410 and 500 K using the GEM instrument at ISIS (and processed with a \textit{Q}\textsubscript{max} of 20 Å$^{-1}$ for a fairer comparison), and previously published in support of persistent OD behaviour at high temperature \cite{Senn2016}. The average structure of BaTiO$_{3}$ is tetragonal at 293 and 350 K and cubic at 410 and 500 K, inviting a comparison of the local structure of BaTiO$_{3}$ at high pressure and high temperature.

At 0.24 and 1.19 GPa consistent improvements in $R_w$ over all $r_\textup{max}$ (shown in Figure \ref{f4}) indicate that the local and medium-range structure of BaTiO$_{3}$ is best described by a rhombohedral displacement of the Ti atom. The refined $|$Q(Γ$_4^-$)$|$ values for an (a,0,0) OPD are approximately $\sqrt{3}/2$ smaller than those of an (a,a,a) OPD suggesting that we are essentially resolving a projection of the [111] type displacement onto the [100] direction. 

The results for the local structure of BaTiO$_{3}$ at 1.19 GPa are essentially very similar to those found for the structure at 0.24 GPa and a decrease in $|$Q(Γ$_4^-$)$|$ of ca.~15 \% points towards a small pressure-induced hardening of the local potential describing the off-centre displacements. These results are comparable with those of the variable temperature PDFs, measured at 293 and 350 K. $R_w$ values at 293 and 350 K again favour the rhombohedral-type displacements up to $r_\textup{max}$ = 10 \angstrom, after which, fitting statistics favour the tetragonal model, indicating sensitivity of the PDF to the average, long-range structure. $|$Q(Γ$_4^-$)$|$ values are in good agreement with the variable pressure results. Again, relative $|$Q(Γ$_4^-$)$|$ values for (a,0,0) compared to (a,a,a) OPDs suggest the resolution of the [111] type displacement onto the [100] direction.

Results for the local structure of high pressure cubic BaTiO$_{3}$ point toward a departure from the local structure behaviour of the high pressure tetragonal structure, and perhaps more interestingly, from the local structure of the high temperature cubic structure. At 2.55 and 4.18 GPa, $|$Q(Γ$_4^-$)$|$ at $r_{\textup{max}}$ = 4 \angstrom~becomes suppressed by ca.~1/2 (cf. 0.24 GPa) and the magnitudes of $|$Q(Γ$_4^-$)$|$ with OPD (a,0,0) and (a,a,a) are approximately equal. Over all $r_\textup{max}$ there is negligible difference between the $R_w$ values for models of tetragonal and rhombohedral Ti displacements. At 2.55 GPa the difference between cubic models and models with off centre displacements decreases approximately linearly until $r_{\textup{max}}$ = 20 \angstrom, after which the difference in $R_w$ drops below significance, whereas at 4.18 GPa this occurs at $r_{\textup{max}}$ = 10 \angstrom. At 4.18 GPa, by 16 \angstrom, $|$Q(Γ$_4^-$)$|$ refines to zero, suggesting that the correlation length of the Ti displacements is below four unit cell lengths. The suppression of $|$Q(Γ$_4^-$)$|$, isotropy of the displacement with respect to the different OPD, and the reduction in correlation lengths are all consistent with the ferroelectric instability in BaTiO$_{3}$ being well-described by the harmonic approximation at elevated pressures. 

On the other hand, our results against previously-published high temperature PDF data clearly favour an (a,a,a) OPD, consistent with the model of chains of rhombohedrally displaced off-centre Ti atoms, which retain substantial correlations along 〈100〉 directions. At 410 and 500 K, refined $|$Q(Γ$_4^-$)$|$ values over $r_\textup{max}$ = 4--10  \angstrom) are similar to those observed at lower temperatures (at $r_{\textup{max}}$ = 4 \angstrom, $|$Q(Γ$_4^-$)$|$ at 293 K = 0.094 \angstrom, 350 K = 0.074 \angstrom, 410 K = 0.071 \angstrom, 500 K = 0.071 \angstrom), but drop to values that are ca.~2/3 of those observed over longer $r_\textup{max}$. The persisting sensitivity to off-center displacements in the high temperature cubic regime is consistent with the model of correlated chains of [111] displacements projected along the [100] axis and lends further support to the OD model for the temperature-induced phase transition.

We find that our observed high pressure trends of the local structure agree with the work of Ravy et al.\cite{Ravy2007} who report diminishing diffuse scattering planes at high pressure and broadening of the diffuse features indicative of a decrease in correlation length of Ti chains, which they discuss in the wider context of pressure-induced Ti-centering. Correlation lengths of ca.~six unit cell lengths (ca.~24 \angstrom) implied by broadened diffuse features at ca.~4 GPa are also in good agreement with our results, where diffuse scattering is sensitive to chain correlations and the PDF method will average chain and non-chain interactions. While the reported diffuse scattering is sensitive to chain correlations, it is less sensitive to the precise nature of the local symmetry breaking.  On the other hand, the method we report here for analysing our high-pressure PDFs has a higher degree of sensitivity to the local symmetry breaking at low $r_{\textup{max}}$ ,but will average over chain and non-chain interactions at high $r_{\textup{max}}$, and thus the two approaches should be viewed as providing complimentary information.  

We stress that although XAS measurements suggest continual off-centre Ti displacements up to 10 GPa\cite{Itie2006}, the sensitivity of the technique is limited to the immediate local environment of the probe atom, extending as far as the next-nearest neighbour only. This makes it difficult to judge how these results differ from those expected from the root mean square displacement of a harmonic oscillator---estimated to be 0.05 \angstrom~at 4.18 GPa from our $r_{\textup{max}}$ = 4 \angstrom~refinements (see Figure \ref{f4}). 

Our results not only show robustly that the OD behaviour of BaTiO$_{3}$ is suppressed at high pressure, but also adds to an emerging research direction on neutron local structure measurements of crystalline materials under hydrostatic pressure\cite{Playford2017,Herlihy2021}, where local structure analysis approaches such as the symmetry motivated approach we have used here can be applied. Such experiments would provide fundamental insight into the pressure induced mode softening in negative thermal expansion materials like ScF${_3}$\cite{Greve2010,Bird2020}, for example, or pressure-induced phase behaviour of framework materials such as Prussian blue analogues\cite{Chapman2006,Bostrom2021}.

\section{Conclusion}
Although it might be tempting to conclude from the average structures that the high temperature and high pressure tetragonal and cubic phases behave in an analogous way, in terms of the local structure, our detailed high pressure PDF study shows that this is not the case. Our symmetry motivated approach of interrogating the local structure of BaTiO$_{3}$ reveals that at high pressure, the OD model provides a less satisfactory description. By 2.55 GPa already, significant suppression of the mode amplitude over short $r_\textup{max}$, isotropy of the OPD and loss of sensitivity to correlated Ti displacements at high pressure all point towards a more harmonic character of the polar mode, which contrasts the high temperature behaviour.

\section*{Acknowledgements}
We thank Professor David Keen for supplying the total scattering data from GEM. A. H. thanks the Science and Technology Facilities Council and the University of Warwick for a studentship. M. S. S. acknowledges the Royal Society for a University Research Fellowship (UF160265) and the EPSRC for funding (EP/S027106/1). We are grateful to STFC for the provision of neutron beam time at ISIS, supported under experiment number RB1910162\cite{DATA}.

\appendix
\section{Validation of PDF sensitivity}

\begin{figure}[t]
	\includegraphics{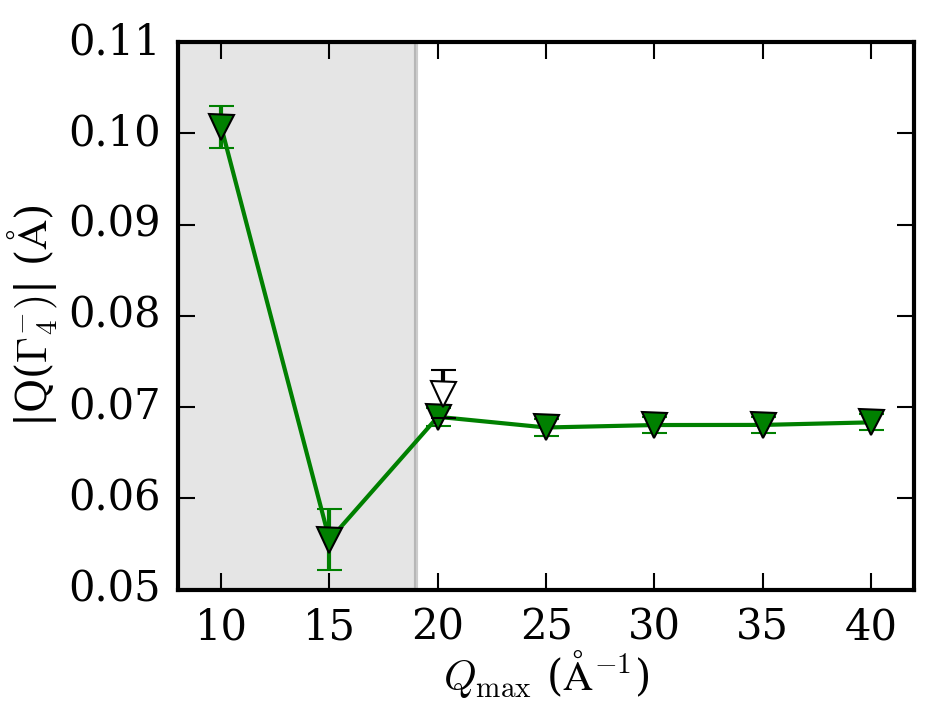}
	\caption{\label{f2} Refined $|$Q(Γ$_4^-$)$|$ values for rhombohedrally constrained Ti and O distortions for a room temperature BaTiO$_{3}$ PDF measured on GEM and processed with \textit{Q}\textsubscript{max} values ranging from 10--40 Å$^{-1}$ in steps of 5 Å$^{-1}$ (green filled markers). The unfilled marker represents the $|$Q(Γ$_4^-$)$|$ value for the lowest pressure (0.24(2) GPa), ambient temperature measurement of BaTiO$_{3}$ on PEARL.}
\end{figure}

\begin{figure}[t]
	\includegraphics{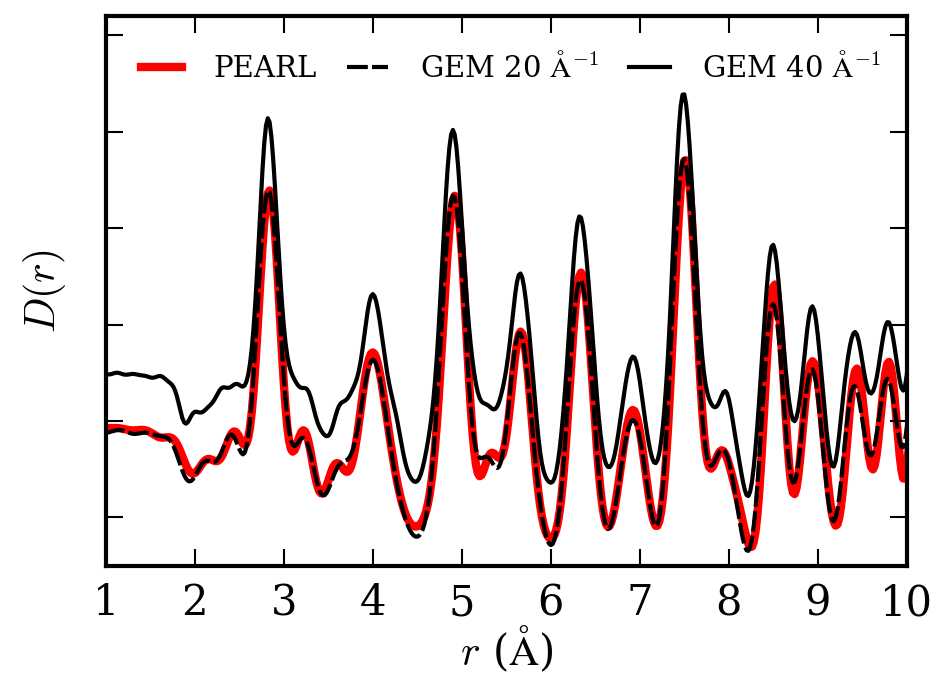}
	\caption{\label{f3} PDFs of BaTiO$_{3}$ measured on PEARL at 0.24 GPa, processed with a \textit{Q}\textsubscript{max} of 20.32 Å$^{-1}$, and on GEM at ambient pressure and temperature, processed with \textit{Q}\textsubscript{max} values of 20 and 40 Å$^{-1}$ (offset in the y-direction).}
\end{figure}

In order to evaluate the sensitivity of our modelling approach to limited \textit{Q}\textsubscript{max} values of the PEARL data, PDFs were processed from room temperature \textit{S}(\textit{Q})s measured on GEM, from reference \citenum{Senn2016} with artificially lowered \textit{Q}\textsubscript{max} values of 10--40 Å$^{-1}$ in steps of 5 Å$^{-1}$. A rhombohedral model was fit to this contiguous series of PDFs, over the $r$-range 1.2--10 Å and the resulting $|$Q(Γ$_4^-$)$|$ values for Ti are shown in Figure \ref{f2}. They show notable consistency over the range 20--40 Å$^{-1}$ and crucially, the values refined against our 0.24 GPa PEARL data with \textit{Q}\textsubscript{max} = 20.32 Å$^{-1}$ are in excellent agreement, falling within error of each other. At \textit{Q}\textsubscript{max} = 15 Å$^{-1}$ the values become inconsistent with those found for higher resolution PDFs. This establishes the lower limit in \textit{Q}\textsubscript{max} with respect to extracting physically meaningful displacements in this reported study and validates the sensitivity of the high pressure PDFs with respect to the local order parameters that we seek to probe. A 293 K total scattering measurement of BaTiO$_{3}$ on the GEM instrument provides a good comparison to the lowest pressure (0.24(2) GPa), ambient temperature PEARL measurement. Figure \ref{f3} shows that PDFs measured on GEM and PEARL (i.e.~in a vanadium can and PE press respectively) and processed with the same  \textit{Q}\textsubscript{max} very closely agree, again, validating the high-pressure data.

\bibliographystyle{apsrev4-1}
\bibliography{prb_batio3highpressure.bib}

\end{document}